\documentclass[smallcondensed]{svjour3}
\usepackage{balance}
\usepackage{amsmath,amssymb,amsfonts}
\usepackage{algorithmic}
\usepackage{graphicx}
\usepackage{svg}
\usepackage{textcomp}
\usepackage{hyperref}
\usepackage{fancyvrb}
\usepackage{url}
\usepackage{xcolor,colortbl}
\usepackage{todonotes}
\usepackage{framed}
\usepackage{siunitx}
\usepackage{multirow}
\usepackage{natbib}
\usepackage{enumitem}
\usepackage[title]{appendix}
\makeatletter
\def\input@path{{.}{reports/}}
\makeatother
\graphicspath{{.}{reports/}} 

\definecolor{Gray}{gray}{0.85}
\definecolor{LightGray}{gray}{0.95}
\definecolor{DarkGray}{gray}{0.65}

\renewcommand*\descriptionlabel[1]{\hspace\labelsep
  \normalfont\bfseries #1:}

\newcommand{\energyPattern}[3]{
  \begin{description}
    \item[Context] #1
    \item[Solution] #2
    \item[Example] #3
  \end{description}
}

\newcommand{\rqn}[1]{\textbf{RQ{#1}:}}
\newcommand{\rquestion}[2]{\begin{framed}%
  \noindent\rqn{#1} #2
\end{framed}}
\definecolor{darkgreen}{RGB}{40,160,40}
\begin{document}
\title{Catalog of Energy Patterns for Mobile Applications}

\author{Luis Cruz \and Rui Abreu}
\institute{Luis Cruz
    \at INESC ID and University of Porto, Portugal\\\email{luiscruz@fe.up.pt}
    \and
    Rui Abreu
    \at INESC ID and IST, University of Lisbon, Portugal\\\email{rui@computer.org}
}

\maketitle

\begin{abstract} Software engineers make use of design patterns for reasons
that range from performance to code comprehensibility. Several design patterns
capturing the body of knowledge of best practices have been proposed in the
past, namely creational, structural and behavioral patterns. However, with the
advent of mobile devices, it becomes a necessity a catalog of design patterns
for energy efficiency. In this work, we inspect commits, issues and pull
requests of 1027 Android and 756 iOS apps to identify common practices when
improving energy efficiency. This analysis yielded a catalog, available online,
with 22 design patterns related to improving the energy efficiency of mobile apps. We
argue that this catalog might be of relevance to other domains such as
Cyber-Physical Systems and Internet of Things. As a side contribution, an
analysis of the differences between Android and iOS devices shows that the
Android community is more energy-aware.

\end{abstract}


\section{Introduction}

The importance of providing developers with more knowledge on how they can
modify mobile apps to improve energy efficiency has been reported in previous
works~\citep{li2014investigation,robillard2016disseminating}.
In particular, mobile apps often have energy requirements
but developers are unaware that energy-specific design patterns do
exist~\citep{manotas2016empirical}. Moreover, developers have to support multiple
platforms while providing a similar user experience~\citep{an2018automatic}.

Design patterns have been formalized to provide general, reusable
solutions to recurrent problems in software design. According to their main
purpose, design patterns were originally categorized in \emph{creational},
\emph{structural}, and \emph{behavioral} patterns~\citep{vlissides1995design}.
Further efforts have leveraged domain-specific catalogs of design patterns to
meet non-functional requirements such as security~\citep{yskout2015security}.
Design patterns also play an important role in educating developers, since
they tend to learn by looking at code examples or using boilerplate code
following well defined solutions~\citep{pham2015automatically,pham2013creating}.

There is a number of practices from experienced developers that lie in the
history of mobile app projects~\citep{negara2014mining,palomba2018community}.
In this work, we collect the set of patterns that developers adopt to improve the
energy efficiency of their apps. We analyze $1783$ apps from Android ($1027$)
and iOS ($756$) and compare practices of developers towards energy
efficiency amongst the two most popular mobile platforms.

In particular, we aim to answer the following questions:

\rquestion{1}{Which design patterns do mobile app developers adopt to improve
energy efficiency?}

We describe a catalog of 22 patterns that mobile app developers resort to when
addressing energy efficiency. We document these patterns so that other
developers learn more about energy best practices and reuse them in their
projects.

\rquestion{2}{How different are mobile app practices addressing energy
efficiency across different platforms?}

We found that Android developers have higher awareness towards energy
efficiency improvement than iOS developers. We show the prevalence of each
energy pattern in the two platforms and discuss potential causes.

This paper makes the following contributions:

\begin{itemize}

  \item We propose a catalog of energy patterns with a detailed description and instructions
  for mobile app developers and designers. It is available
  online:~\url{https://tqrg.github.io/energy-patterns}, and we welcome
  contributions from the community as pull request.

  \item We provide a dataset with $1563$ commits, issues, and pull requests in which mobile
  app development practitioners address the energy efficiency of their apps. The
  dataset and collection tools are available
  online:~\url{https://github.com/TQRG/energy-patterns}.

  \item We compare energy efficiency awareness in mobile app development in
  different platforms (viz. Android and iOS).

\end{itemize}

The remainder of this paper is organized as follows. Related work is discussed
in Section~\ref{sec:rw}. Section~\ref{sec:methodology} outlines the methodology
used to collect data and extract energy patterns in our study, followed by
Section~\ref{sec:energy_patterns} describing the collection of energy patterns.
Section~\ref{sec:data_summary} summarizes the collected data and discusses
implications of the list of proposed patterns. Threats to the validity are
discussed in Section~\ref{sec:t2v}. Finally, we draw our conclusions and point
directions for future work in Section~\ref{sec:conclusions}.
\section{Related Work}
\label{sec:rw}

Improving energy efficiency of mobile apps has gained the attention of the
research community recently, which addressed the challenge in different ways:
identifying energy
bugs~\citep{pathak2011bootstrapping,banerjee2014detecting,vekris2012towards},
profiling energy
consumption~\citep{wilke2013jouleunit,liu2013has,hao2013estimating,behrouz2015ecodroid,pathak2011fine,zhang2010accurate,chowdhury2018greenscaler,di2017software,hindle2014greenminer,romansky2017deep},
or understanding best coding practices for energy
efficiency~\citep{sahin2016benchmarks,cruz2017performance,cruz2018using,linares2014mining,pathak2012energy}.

In previous work, \cite{moura2015mining} have mined 290 energy-saving software
commits, identifying 12 categories of source code modification to improve energy
usage~\citep{moura2015mining}: \emph{Frequency and voltage scaling}, \emph{Use
power efficient library/device}, \emph{Disabling features or devices},
\emph{Energy bug fix}, \emph{Low power idling}, \emph{Timing out}, \emph{Avoid
polling}, \emph{Pin management}, \emph{Display and UI tuning}, \emph{Avoid
unnecessary work}, \emph{Miscellaneous}, and \emph{Outlier}. The programming
languages used to implement the software systems used in this study were
diverse: programming C (158 projects), Java (25 projects), Bourne Shell (17
projects), Arduino Sketch (15 projects), and C++ (12 projects). They found that
roughly 50\% of energy-saving commits target lower levels of the software stack
(e.g., kernels and drivers), which is not a level of abstraction commonly
considered during the design of mobile apps. Our work extends this approach to
the ecosystem of mobile apps by compiling a set of coding practices that can be
used by practitioners across mobile apps on different platforms. Thus, our
dataset of apps also includes projects written in \emph{Swift},
\emph{Objective-C}, \emph{Java}, \emph{Kotlin}, and any other language used for
mobile app development in iOS or Android. In addition, we detail these and
other energy-saving categories with a context and guidelines to help
developers decide on the most appropriate pattern. Moreover, we compare the
prevalence of these patterns across different mobile platforms.

With a similar approach, \cite{bao2016android} have mined 468 power management
commits to find coding practices in Android apps~\citep{bao2016android}. Using
a hybrid card sort approach, six different power management practices were
identified: \emph{Power Adaptation}, \emph{Power Consumption Improvement},
\emph{Power Usage Monitoring}, \emph{Optimizing Wake Lock}, \emph{Adding Wake
Lock and Bug Fix \& Code Refinement}. The study shows that power management
activities are more prevalent in navigation apps. Conversely, our work
focuses on energy-saving commits, pull requests, and issues. Using the same
taxonomy, our work concentrates exclusively on coding practices for \emph{Power
Adaptation}, and \emph{Power Consumption Improvement}. Moreover, rather than
analyzing the prevalence of power management activities amongst different app
categories, we emphasize on providing actionable findings for mobile app
practitioners. Finally, we extend this work to the iOS mobile platform, which
shares a big part of the mobile app market.

Previous work studied the views of mobile app developers on energy
efficiency improvement by mining
\emph{StackOverflow}\footnote{\emph{StackOverflow} is a collaborative Web
platform for questions and answers on a wide range of topics in computer
programming.} posts~\citep{pinto2014mining}. It was found that developers make
interesting questions about energy efficiency problems. However, the answers
provided on this topic are often flawed or vague. Our work analyses mobile app
projects to collect recurrent solutions adopted by developers.

There is work that studied the impact of performance optimizations on the energy
efficiency of mobile
apps~\citep{hecht2016empirical,cruz2017performance,linares2014mining,sahin2016benchmarks}.
However, only platform-specific optimizations were addressed ---
e.g., in early versions of Android using \texttt{get}/\texttt{set} methods
internally was less energy efficient than accessing fields directly. In our
work, we focus on patterns that can be applied regardless of the mobile
platform being used.

Furthermore, related work studied the impact of high-level coding and design
practices. E.g., the use of advertisement increases energy usage of mobile
apps~\citep{pathak2012energy}, bundling small HTTP requests can be used to
enhance energy efficiency~\citep{li2014investigation}. We assess how mobile app
developers are using these and other patterns to improve energy efficiency in
real mobile apps. Measuring the effective impact of these optimizations on
energy efficiency is out of the scope of our work.

Previous work has delivered a catalog of quality smell
patterns for Android apps~\citep{reimann2014tool}. Our work differentiates by
focusing on energy
efficiency improvements and including iOS apps. Still, there is one pattern
that is common in the two catalogs: Reimann et al.'s \emph{Early Resource Binding} and
our's \emph{Open Only When Necessary}.

The impact of general purpose software design patterns on energy efficiency has
been studied in previous work~\citep{sahin2012initial}. It was shown that
design patterns may have different impacts on energy consumption. Related work
has also evaluated the impact of different machine learning
algorithms~\citep{mcintosh2018can}. The most efficient technique algorithm
depends on properties such as the size of the dataset, and the number of data
attributes. Our work differs as we exclusively focus on patterns that
are applied to improve the energy efficiency of mobile apps.

Most of the works described above focus on a single platform, notably Android.
As recently reported by the Google Android VP Dave Burke, Google may actually
have less than 66\% global market share. Hence, we have also considered iOS in
our study. Related work has extended the study of coding practices to iOS. In a
study with $279$ iOS apps and $1551$ Android apps, no significant distinction
was found in the prevalence of code smells between iOS and Android
apps~\citep{oliveira2017study}. A different work has studied error handling
practices of Swift\footnote{Swift is the official programming language for iOS
apps.}~\citep{cassee2018swift}. Nearly half of the $2733$ Swift projects did
not exhibit any error handling code. Our work provides more enlightenment on
practices of developers amongst the iOS ecosystem.

\section{Methodology}
\label{sec:methodology}

\begin{figure}
  \centering
  \includegraphics[width=\textwidth]{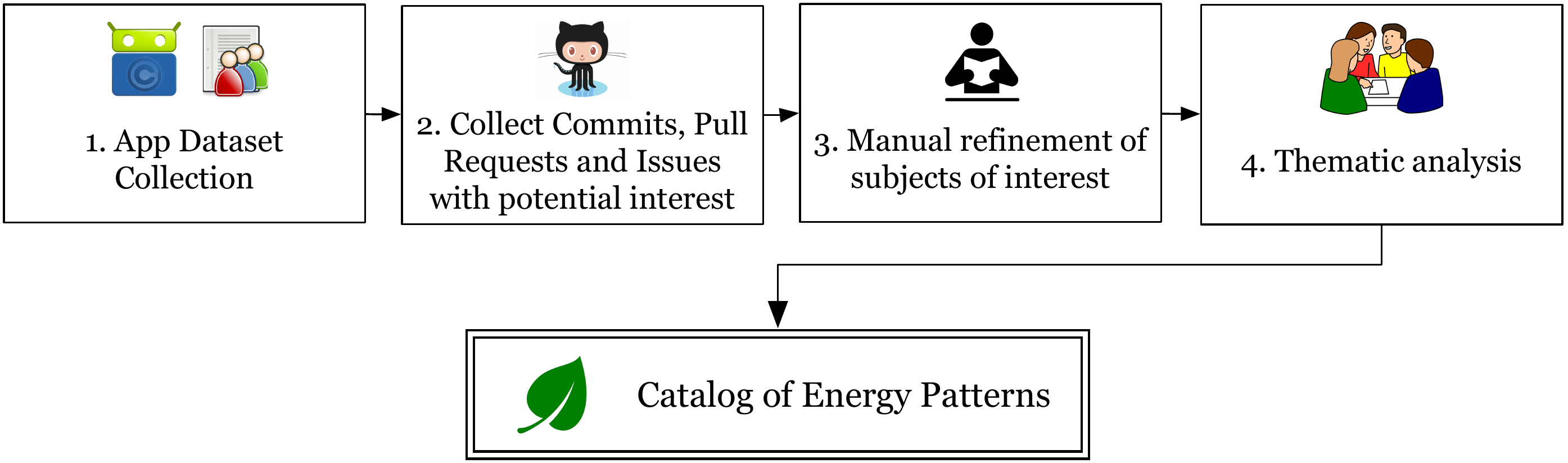}
  \caption{Methodology used to extract energy patterns from mobile apps.}
  \label{fig:data_collection}
\end{figure}

We designed a methodology to extract energy patterns from existing mobile apps.
In total, we collect a total of 1027 Android apps and 756 iOS apps, including
apps designed for smartphones, tablets, wearables, or e-paper devices.
Essential to our analysis, apps from both platforms are open source and have
their git repositories available on GitHub\footnote{GitHub is a social coding
platform with a git web interface.}. Our methodology is illustrated in
Figure~\ref{fig:data_collection} and comprised the following four main tasks:

\begin{itemize}
  \item App dataset collection
  \item Automatic gathering of subjects with potential interest (i.e., commits, issues, and pull requests)
  \item Manual refinement of subjects of interest
  \item Thematic analysis (infer energy patterns) according to the solution encountered to improve
  energy efficiency
\end{itemize}

\subsection{App Dataset Collection}

Multiple open source mobile app catalogs were combined to collect Android and iOS
mobile apps. For Android, we resort to \emph{F-Droid}, a catalog that lists $2800$
free and open source Android apps\footnote{F-droid's website: \url{https://f-droid.org/}
(Visited on \today).}. There are open source apps that are not available in
\emph{F-droid} for not fulfilling free software requirements (e.g., Signal
app\footnote{Signal is an open source messaging app: \url{https://signal.org} (Visited on \today).}).
Although these are just a minority of apps we argue that they can provide
relevant input on energy efficiency practices. Thus, we included Android apps listed
in community-curated collections of Android open source apps\footnote{\emph{Amazing open source
Android apps} curated list available here: \url{https://github.com/Mybridge/amazing-android-apps}
(Visited on \today). }. This resulted in $1027$ apps --- $1001$ from \emph{F-droid}
and 26 from curated lists.

For iOS we use the \emph{Collaborative List of Open-Source iOS Apps}~\footnote{The list is
available here: \url{https://github.com/dkhamsing/open-source-ios-apps} (Visited on \today)},
amounting to
829 apps listed with the help of a community of 195 collaborators. Given our
constraint of having a publicly available GitHub repository, we ended up including
756 iOS apps in our study.

\begin{figure}
  \centering
  \includegraphics[width=0.7\textwidth]{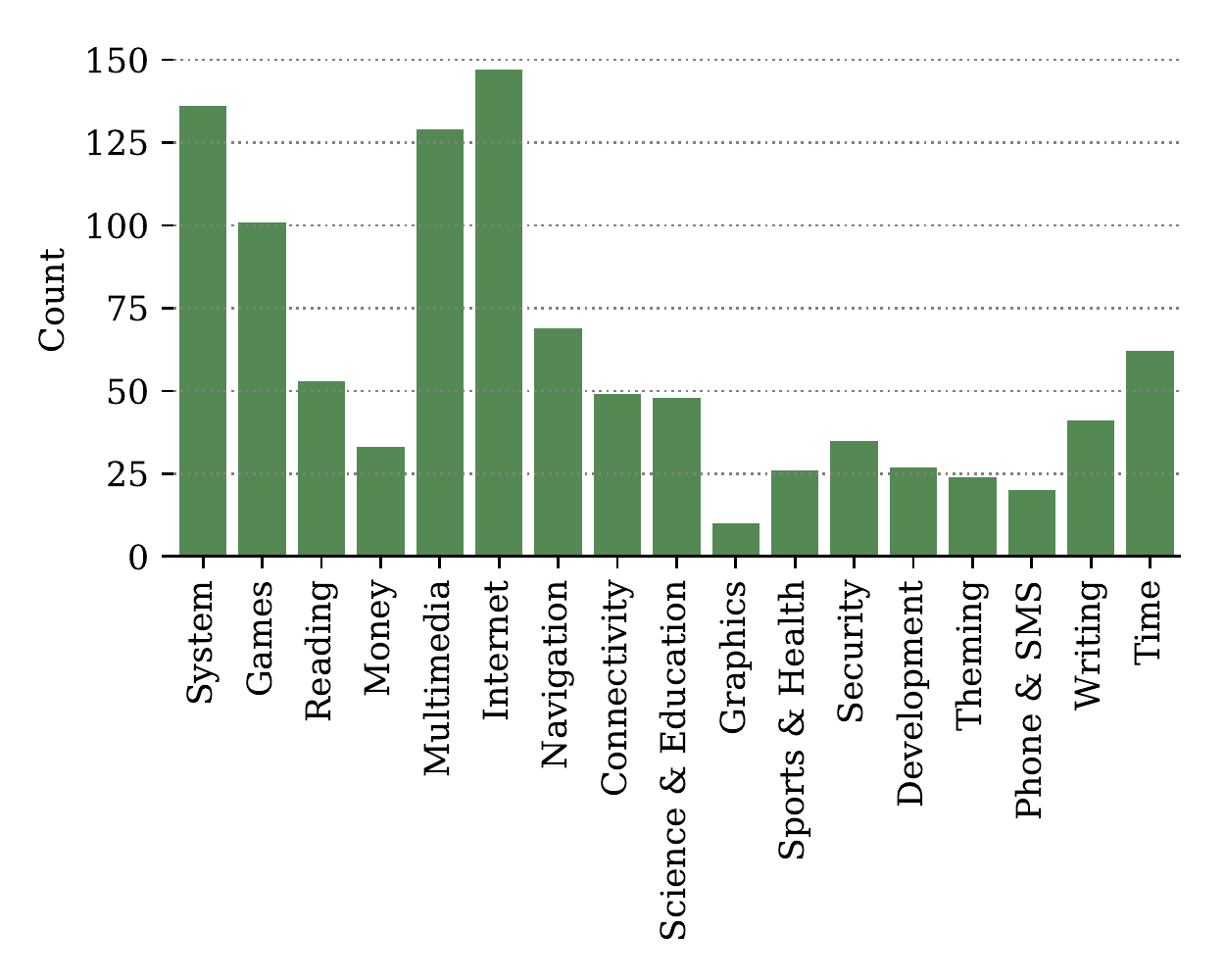}
  \caption{Distribution of categories in Android apps.}
  \label{fig:android_categories}
\end{figure}

\begin{figure}
  \centering
  \includegraphics[width=0.7\textwidth]{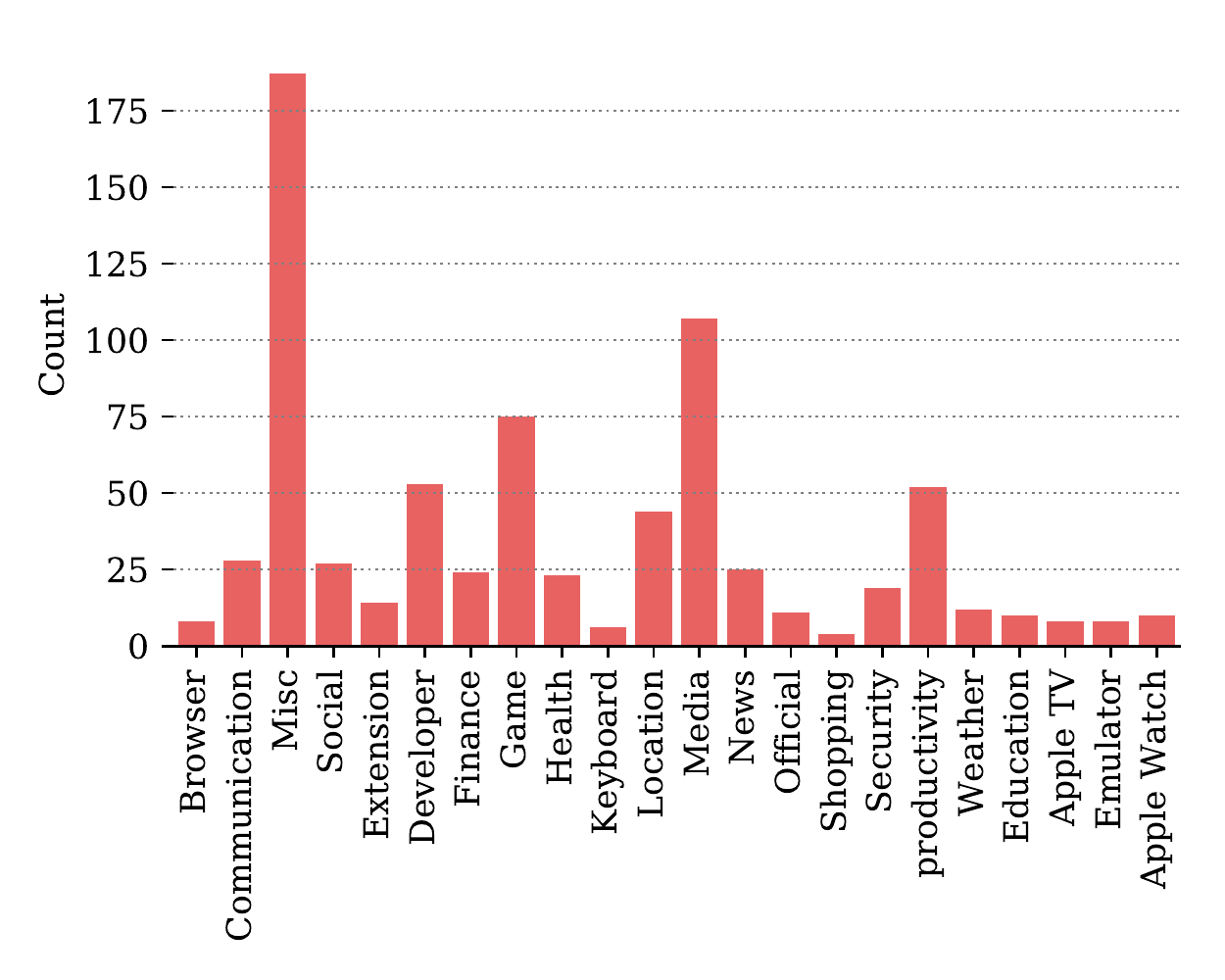}
  \caption{Distribution of categories in iOS apps.}
  \label{fig:ios_categories}
\end{figure}

The apps used in this study are from a wide range of categories for Android and
iOS, as depicted in Figure~\ref{fig:android_categories} and
Figure~\ref{fig:ios_categories}, respectively. In addition,
Table~\ref{tab:github_stars} shows the dispersion of apps in terms of
popularity metrics: GitHub stars, GitHub forks, number of reviews, and rating.
The number of reviews and the ratings are from the apps in the dataset
that are published in the Google Play Store or the iOS App Store. Thus, only
from a subset of the apps in this study: 64\% of Android apps and
20\% of iOS apps. GitHub stars go up to roughly 15K in both platforms, while
GitHub forks up to 8K in Android and 5K in iOS. On average, the apps have a
rating of 4 out of 5.

\begin{table}[htbp]
  \caption{Descriptive statistics of the popularity metrics of the Android and iOS apps in the dataset.}
\begin{center}
  \label{tab:github_stars}
\resizebox{\textwidth}{!}{
\begin{tabular}{llrrrrrrr}
\hline
                                    & Platform   &   Mean &   Std &   Min &   25\% &   Median &   75\% &   Max \\
\hline                              
\multirow{2}{*}{Stars}              & Android    &   145   &    608   &   0   &   6   &     20   &   68   &    15159   \\
                                    & iOS        &   486   &   1272   &   0   &  18   &     71   &  329   &    15318   \\
\hline                              
\multirow{2}{*}{Forks}              & Android    &    75   &    302   &   0   &   4   &     14   &   46   &     7811   \\
                                    & iOS        &   118   &    354   &   0   &   7   &     20   &   71   &     4820   \\
\hline
\multirow{2}{*}{Number of Reviews*}  & Android    & 31855   & 529676   &   1   &  24   &    138   & 1087   & 13080790   \\
                                    & iOS        &  3241   &  12597   &   5   &  18   &     75   & 1038   &   115011   \\
\hline
\multirow{2}{*}{Rating*}             & Android    &     3.8 &      0.5 &   1.0 &   4.0 &      4.0 &    4.0 &        5.0 \\
                                    & iOS        &     4.1 &      0.7 &   2.0 &   3.6 &      4.0 &    4.5 &        5.0 \\
\hline
\multicolumn{9}{l}{\footnotesize *as in Google Play Store and iOS App Store.}\\
\hline
\end{tabular}
}
\end{center}
\end{table}

\subsection{Automatic gathering of commits, issues, and pull requests}

In this step, we collect from all GitHub repositories in our dataset any
commit, issue or pull request that potentially contains energy improvement
practices. As done in previous work~\citep{bao2016android}, any instance that
mentions the words \emph{energy}, \emph{battery}, or \emph{power} is selected.
The following regular expression is used:

\begin{center}
{
\noindent
\fbox{
\texttt{.*(energy|battery|power).*}
}
}
\end{center}

The \emph{GitHub API v3} was used to automatically collect data from public
repositories. Note that we only include commits that were merged in the
\textit{default} branch of the projects. In total, we gathered $6028$ entries
that matched this regular expression.


\subsection{Manual Refinement}

We understand that the regular expression used in the automatic data collection
yields many false positives. As an example, consider the following entries
collected in the previous step:

\begin{itemize}

  \item \emph{\color{red}``Adding a link to the app’s device settings in iOS Settings.app would
    be great for \textbf{\underline{power}} users.''} (False positive found in the app
  \emph{WordPress} for iOS\footnote{The whole thread can be found here:
  \url{https://github.com/wordpress-mobile/WordPress-iOS/issues/6057} (Visited
  on \today).}).

  \item \emph{\color{red}``(\dots) recently a lot of issues that the core team does not
  have the \textbf{\underline{energy}} to implement themselves have been closed.''} (False
  positive found in the app \emph{Minetest} for Android\footnote{The whole thread can
  be found here: \url{https://github.com/minetest/minetest/issues/6394}
  (Visited on \today).}).

  \item \emph{\color{darkgreen}``One thing is really important on mobile devices, and that is \textbf{\underline{power}} consumption''} (True positive found in the app \emph{ChatSecure} for iOS\footnote{The whole thread can
  be found here: \url{https://github.com/ChatSecure/ChatSecure-iOS/issues/31}
  (Visited on \today).}).
\end{itemize}

Although the first two examples match with the regular expression, they are not
expected to deal with energy-related practices. On contrary, the last example is
referring to the topic of power consumption. Thus, it is likely to provide
useful insights on energy improvement practices.

To filter out unrelated entries, we resort to a manual analysis of each
instance, comprising two iterations:

\begin{enumerate}

  \item Check the line where a match with the regular expression was
  found. If the sentence does not mention anything related to energy
  consumption, the subject is discarded from the dataset.

  \item Check the whole thread in which the mention was found. I.e., open the
  GitHub page where the commit, issue, or pull request is documented and
  analyze the context in which the energy topic is being discussed. This step
  removes cases in which contributors are discussing the topic of energy for
  contexts that are not related to energy efficiency improvement. E.g., cases
  were found in which developers were talking about the battery of their own
  laptops, or in which the app actually had a feature in which it displayed the
  status of the battery\footnote{An example can be found here:
  \url{https://github.com/hrydgard/ppsspp/issues/7765} (Visited on \today).}.

\end{enumerate}

After this step, we ended up with a total of $1563$ subjects: $332$ commits,
$1089$ issues, and $142$ pull requests.

\subsection{Thematic Analysis}

We resort to a methodology based on Thematic Analysis~\citep{fereday2006demonstrating}
to derive design patterns from commits, issues and pull requests. Thematic
Analysis is a widely-used qualitative data analysis method, that focuses on
identifying patterned meaning in a dataset. Its hybrid process of deductive and
inductive analysis has been successfully used in previous work to categorize
software commits~\citep{moura2015mining}. We follow a similar approach by
adopting a four-stage process:

\begin{description}

  \item[Familiarization with data] We have carefully read the information provided in
  commits, issues or pull-requests, including comments and descriptions. Relevant
  advice and reasoning are collected and studied using online documentation.

  \item[Generating initial labels] For each commit, issue, and pull request, we
  describe the change in a generic short sentence --- i.e., without resorting
  to specific properties of the app. This process was split into several
  iterations to discuss amongst both authors in order to refine the
  labels.

  \item[Reviewing themes] After having all subjects spread in different labels, we
  discuss and review them to find themes that should be merged or split.
  Some themes were discarded for not being present in a sufficient number of
  subjects. In particular, we filter out themes that did not occur in at least
  three different apps. In addition, we corroborate and legitimate coded themes,
  by finding evidence in the literature that supports or discards themes.

  \item[Defining and naming themes] In this stage, we make a structured
  description of each theme to provide a set of straight-forward guidelines
  that can be reused in the design of different mobile app projects. Each theme
  is now converted into an \textbf{Energy Pattern}. Each pattern includes a
  \emph{name}, a brief \emph{description}, a \emph{context} or problem in which
  the pattern can be applied, and a \emph{solution} with instructions on how to
  apply the pattern. The solutions provided are based on a combination of
  authors' experience, logical arguments, literature, mobile platform
  documentation, and the data itself.

\end{description}

In total 332 commits, 1089 issues, and 142 pull requests were analyzed using this
approach. As a result, we identify and document 22 energy patterns that appear
in 431 of the analyzed subjects.

\subsection{Reproducibility-Oriented Summary}
\label{sec:reproducibility}

Based on previous guidelines for app store analyses~\citep{martin2017survey},
we describe ours as follows to help repeat the experimental procedure and
reproduce the results:

\begin{description}
    \item[App Stores used to gather collections of apps] We use apps
    available on \emph{F-Droid}, and on the list
    \emph{Collaborative List of Open-Source iOS Apps}.

    \item[Total number of apps used] The study comprises $1783$ apps.

    \item[Breakdown of free/paid apps used in the study] Only non-paid apps are
    listed in our dataset.

    \item[Categories used] All categories were included in this study.

    \item[API usage] GitHub REST API v3\footnote{Documentation of GitHub REST
    API v3 available here: \url{https://developer.github.com/v3/} (Visited on \today).}.

    \item[Whether code was needed from apps] Source code was required
    to analyze code changes in commits.

    \item[Fraction of open source apps] Open source apps are used
    exclusively.

    \item[Static analysis techniques] No static analysis was performed. In some
    tasks, the code was analyzed manually by the authors.

\end{description}

All scripts and tools developed in this work are publicly available with an
open source license: \url{https://github.com/TQRG/energy-patterns}.

\section{Energy Patterns}
\label{sec:energy_patterns}

In this section, we present the energy patterns collected in this study. As
mentioned before, each energy pattern is described by the following entries:
context, solution, and an example illustrating a practical usage of the
pattern. All these patterns are also available
online:~\url{https://tqrg.github.io/energy-patterns}. The website also provides
links to the occurrences in the apps, disclosing the discussion performed by
developers and practical examples of the patterns in this catalog.

We summarize the occurrences of these patterns in Table~\ref{tab:occurrences},
presenting their frequency in Android and iOS platforms along with an
indication of their prevalence in related work and grey literature (as listed
in the Appendix~\ref{sec:grey_literature}).

\begin{table}[htbp]
  \caption{Energy patterns' occurrences and related work.}
\begin{center}
  \label{tab:occurrences}
\begin{tabular}{>{\raggedright\arraybackslash}p{0.25\textwidth}rr>{\centering\arraybackslash}p{0.3\textwidth}c}
\hline
\rowcolor{DarkGray}
Pattern                                   & Android & iOS & Related Work & Grey Literature \\
\hline
\rowcolor{Gray}
Dark UI Colors                            & 28      &   2 & ~\citep{agolli2017investigating,linares2017gemma,li2014making,li2015nyx} & -\\
\rowcolor{LightGray}
Dynamic Retry Delay                       & 10      &   2 & - & -\\
\rowcolor{Gray}
Avoid Extraneous Work                     & 32      &   9 & - & [\ref{grey:apple_graphics}]\\
\rowcolor{LightGray}
Race-to-idle                              & 27      &   5 & \citep{liu2016understanding,banerjee2016automated,cruz2017performance,pathak2012keeping} &  - \\
\rowcolor{Gray}
Open Only When Necessary                  &  4      &   3 & \citep{banerjee2016automated,reimann2014tool} & -\\
\rowcolor{LightGray}
Push over Poll                            & 13      &   3 & - & [\ref{grey:xpp}, \ref{grey:alternative_push}]\\
\rowcolor{Gray}
Power Save Mode                           & 24      &   5 & - & [\ref{grey:apple_power_save}]\\
\rowcolor{LightGray}
Power Awareness                           & 35      &   6 & \citep{bao2016android} & [\ref{grey:apple_power_save}]\\
\rowcolor{Gray}
Reduce Size                               &  3      &   0 & \citep{boonkrong2015reducing} & [\ref{grey:gzip}] \\
\rowcolor{LightGray}
WiFi over Cellular                        & 13      &   2 & \citep{metri2012eating} & [\ref{grey:wifi_cellular}, \ref{grey:android_connectivity}, \ref{grey:apple_network}] \\
\rowcolor{Gray}
Suppress Logs                             &  7      &   1 & \citep{chowdhury2018exploratory} & - \\
\rowcolor{LightGray}
Batch Operations                          & 17      &   1 & \citep{li2014investigation,corral2015energy,cai2015delaydroid} &  [\ref{grey:android_doze}, \ref{grey:android_battery}, \ref{grey:android_youtube}]\\
\rowcolor{Gray}
Cache                                     & 14      &   4 & \citep{gottschalk2014saving} & [\ref{grey:android_battery}] \\
\rowcolor{LightGray}
Decrease Rate                             & 27      &  10 & - & - \\
\rowcolor{Gray}
User Knows Best                           & 33      &  11 & - & - \\
\rowcolor{LightGray}
Inform Users                              &  6      &   4 & - & - \\
\rowcolor{Gray}
Enough Resolution                         & 10      &   7 & - & - \\
\rowcolor{LightGray}
Sensor Fusion                             & 12      &   3 & \citep{shafer2010movement} & [\ref{grey:apple_core}]\\
\rowcolor{Gray}
Kill Abnormal Tasks                       & 11      &   1 & - & - \\
\rowcolor{LightGray}
No Screen Interaction                     &  8      &   2 & - & - \\
\rowcolor{Gray}
Avoid Extraneous Graphics and Animations  & 11      &   8 & \citep{kim2016content} & [\ref{grey:apple_graphics}]\\
\rowcolor{LightGray}
Manual Sync, On Demand                    &  4      &   5 & - & - \\
\hline
\end{tabular}
\end{center}
\end{table}

\subsection{Dark UI Colors}

Provide a dark UI color theme to save battery on devices with AMOLED\footnote{AMOLED is a display
technology used in mobile devices and stands for Active Matrix Organic Light Emitting Diodes.}
screens~\citep{agolli2017investigating,linares2017gemma,li2014making,li2015nyx}.

\begin{figure}[h]
  \centering
  \includegraphics[width=0.4\textwidth]{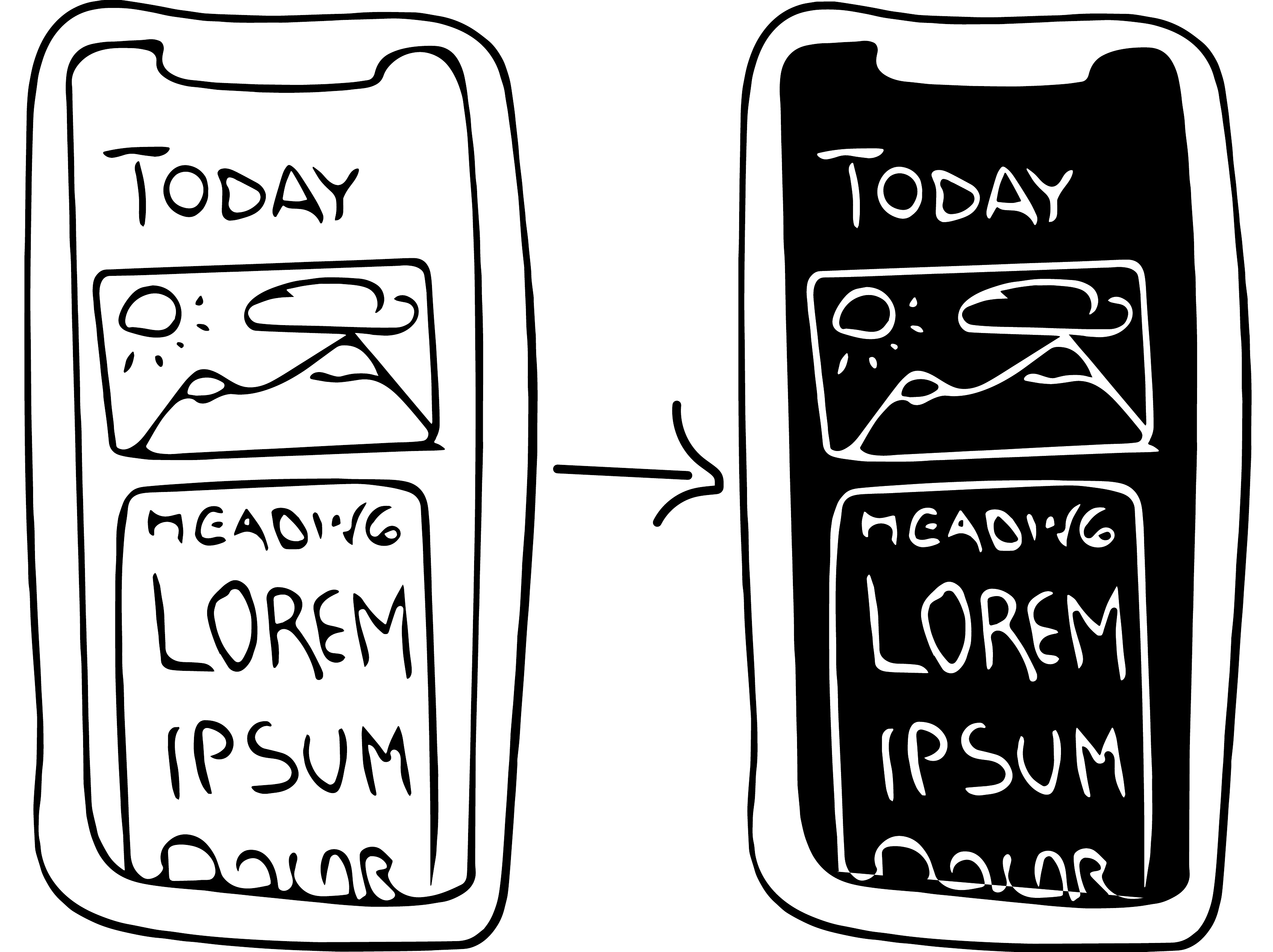}
  \caption{UI themes with dark colors are more energy efficient.}
  \label{fig:dark_ui_colors}
\end{figure}

\energyPattern{

  Screen is one of the major sources of power consumption in mobile devices. Apps
  that require heavy usage of screen (e.g., reading apps) can have a substantial
  negative impact on battery life.

}{

  Provide a UI with dark background colors, as illustrated in
  Figure~\ref{fig:dark_ui_colors}. This is particularly beneficial for mobile
  devices with AMOLED screens, which are more energy efficient when displaying
  dark colors. In some cases, it might be reasonable to allow users to choose
  between a light and a dark theme. The dark theme can also be activated using
  a special trigger (e.g., when battery is running low).

}{

In a reading app, provide a theme with a dark background using light colors
to display text. When compared to themes using light backgrounds, a dark
background will have a higher number of dark pixels.

}

\subsection{Dynamic Retry Delay}

Whenever an attempt to access a resource fails, increase the
time interval before retrying to access the same resource.

\energyPattern{
  Mobile apps that need to collect or send data from/to other resources (e.g.,
  update information from a server). Commonly, when
  the resource is unavailable, the app will unnecessarily try to connect the
  resource for a number of times, leading to unnecessary power consumption.
}{
  Increase retry interval after each failed connection. It can be either a
  linear or exponential growth. Update interval can be reset upon a successful
  connection or a given change in the context (e.g., network status).
}{

  Consider a mobile app that provides a news feed and the app is not able to reach
  the server to collect updates. Instead of continuously polling the server
  until the server is available, use the Fibonacci series\footnote{Fibonacci
  series is a sequence of numbers in which each number is the sum of the two
  preceding numbers (e.g., 1, 1, 2, 3, 5, 8, etc.). } to increase the time
  between attempts.

}

\subsection{Avoid Extraneous Work}

Avoid performing tasks that are either not visible, do not have a direct impact
on the user experience to the user or quickly become obsolete. This has been
documented in the iOS online documentation\footnote{\emph{Energy Efficiency
Guide for iOS Apps – Avoid Extraneous Graphics and Animations} available here:
\url{https://developer.apple.com/library/archive/documentation/Performance/Conceptual/EnergyGuide-iO
S/AvoidExtraneousGraphicsAndAnimations.html} (Visited on \today).}.

\energyPattern{
  Mobile apps have to perform a number of tasks simultaneously. There are cases in
  which the result of those tasks is not visible (e.g., the UI is presenting other
  pieces of information), or the result is not necessarily relevant to the user.
  This is particularly critical when apps go to the background. Since the data
  quickly becomes obsolete, the phone is using resources unnecessarily.
}{
  Select a concise set of data that should be presented to the user and
  enable/disable update and processing tasks depending on their effect on the data that is visible or
  valuable to the user.
}{
  Consider a time series plot that displays real-time data.
  The plot needs to be constantly updated with the incoming stream of  data ---
  however, if the user scrolls up/down in the UI view making the plot hidden, the
  app should cease drawing operations related with the plot.
}

\subsection{Race-to-idle}

Release resources or services as soon as possible (such as wake locks,
screen)~\citep{liu2016understanding,banerjee2016automated,cruz2017performance,pathak2012keeping}.

\energyPattern{

  Mobile apps use a number of resources that can be manually closed after use.
  While active, these resources are ready to respond to requests from the app and
  require extra power consumption.

}{

  Make sure resources are inactive when they are not necessary by manually
  closing them.

}{

  Implement handlers for events that are fired when the app
  goes to background, and release wake locks accordingly.

}

\subsection{Open Only When Necessary}

Open/start resources/services only when they are strictly necessary.

\energyPattern{

  Some resources require to be opened before use. It might be tempting to open the necessary
  resources at the beginning of some task (e.g., upon the creation of an activity). However,
  resources will be actively waiting for requests, and consequently consuming energy.

}{

  Open resources only when necessary. This also avoids activating resources that will never be
  used~\citep{banerjee2016automated}.

}{

In a mobile app for video calls, only start capturing video at the moment that
it will be displayed to the user\footnote{This is a real example that can be
found here:
\url{https://github.com/signalapp/Signal-Android/commit/cb9f225f5962d399f48b65d5
f855e11f146cbbcb} (Visited on \today).}
.

}

\subsection{Push over Poll}

Use push notifications to receive updates from resources, instead of actively querying resources
(i.e., polling).

\energyPattern{

Mobile apps need to get updates from resources (e.g., from a server). One way of checking for updates
is by periodically query those resources. However, this will lead to several requests that
will return no update, leading to unnecessary energy consumption.

}{

Use push notifications to get updates from resources. Note -- this is a big challenge amongst FOSS
apps since there is no good open source alternative for Firebase Cloud Messaging (former Google
Cloud Messaging).

}{

In a messaging app, instead of actively check for new messages, the app can subscribe push
notifications.

}

\subsection{Power Save Mode}

Provide an energy efficient mode in which user experience can drop for the sake of better energy usage.

\begin{figure}[h]
  \centering
  \includegraphics[width=0.2\textwidth]{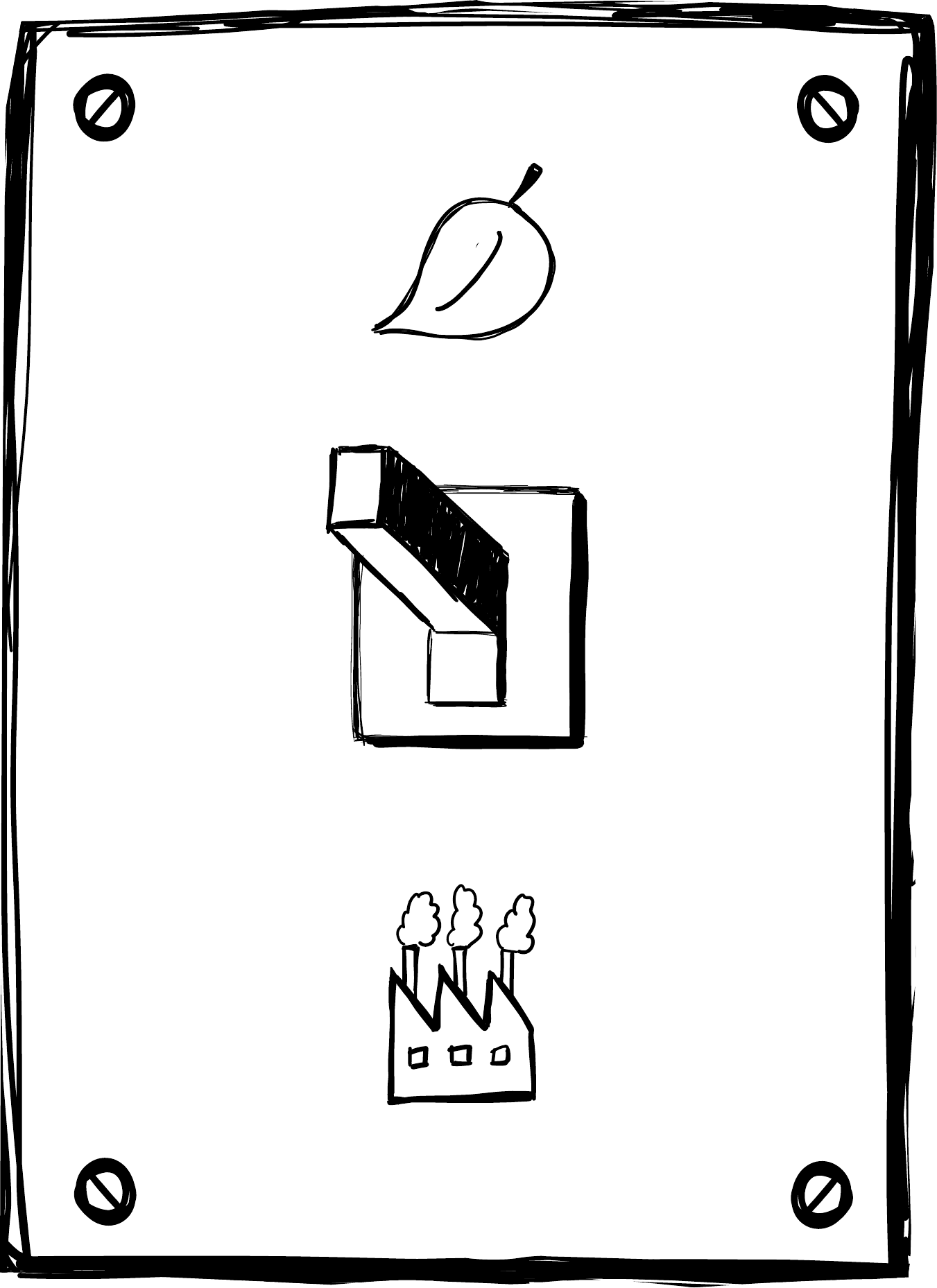}
  \caption{Power Save Mode allows to run the app in two different modes: a fully-featured mode and an energy efficient mode.}
  \label{fig:power_save_mode}
\end{figure}

\energyPattern{

Whenever the device battery is running low, users want to avoid losing connectivity before they
reach a power charging station. If the device shuts down, users might miss important calls or will
not be able to do an important task. Still, apps might be running unimportant tasks that will
reduce battery life in this critical context.

}{

The app provides a power save mode in which it uses fewer resources while providing the minimum
functionality that is indispensable to the user. It can be activated manually or upon some power
events (e.g., when battery reaches a given level). User experience can drop for the sake of energy
efficiency. Note, this is enforced in iOS for some use cases if the apps use the BackgroundSync
APIs.

}{

Deactivate features, reduce update intervals, or deactivate animated effects in the UI.

}

\subsection{Power Awareness}

Have a different behavior when the device is connected/disconnected to a power station or has
different battery levels.

\energyPattern{

There are some features that are not strictly necessary to users although they improve user
experience (e.g., UI animations). Moreover, there are operations that do not have high priority and
do not need to execute immediately (e.g., backup data in the cloud).

}{

Enable or disable tasks or features according to power status. Even when the device is connected
to power, the battery might still be running low, it might be advisable to wait until a pre-defined
battery level is reached (or the power save mode is deactivated).

}{

Delay intensive operations such as cloud syncing or image processing until the device is connected
to a charger.

}

\subsection{Reduce Size}

When transmitting data, reduce its size as much as possible.

\energyPattern{

Data transmission is a common operation in mobile apps. However, such operations
are energy greedy and the time of transmission should be reduced as much as
possible.

}{

Exchange only what is strictly necessary, avoiding sending unnecessary data. Use
data compression when possible.

}{

When performing HTTP requests, use gzip content encoding to compress data.

}

\subsection{WiFi over Cellular}

Delay or disable heavy data connections until the device is connected to a WiFi
network.

\energyPattern{

Data needs to be synchronized with a server but it is not urgent and can be postponed.

}{

Data connections using cellular networks are usually more battery intensive than
connections using WiFi~\citep{metri2012eating}. Low priority operations that require a data connection
to exchange considerable amounts of data should be delayed until a WiFi
connection is available.

}{

Consider a mobile app to organize photos that allows users to backup their photos in a
cloud server. Use an API to check the availability of a WiFi connection and postpone cloud
synchronizing in case it cannot be reached.

}

\subsection{Suppress Logs}

Avoid using intensive logging. Previous work has found that logging activity at
rates above one message per second significantly reduces energy
efficiency~\citep{chowdhury2018exploratory}.

\energyPattern{

Developers resort to logging in their mobile apps to ensure their correct
behavior and simplify bug reporting. However, logging operations create
overhead on energy consumption without creating value to the end user.

}{

Avoid using intensive logging, keeping rates below one message per second.

}{

Disable logging when processing real-time data. If necessary enable only during
debugging executions.

}

\subsection{Batch Operations}

Batch multiple operations, instead of putting the device into an active state many times.

\begin{figure}[h]
  \centering
  \includegraphics[width=0.9\textwidth]{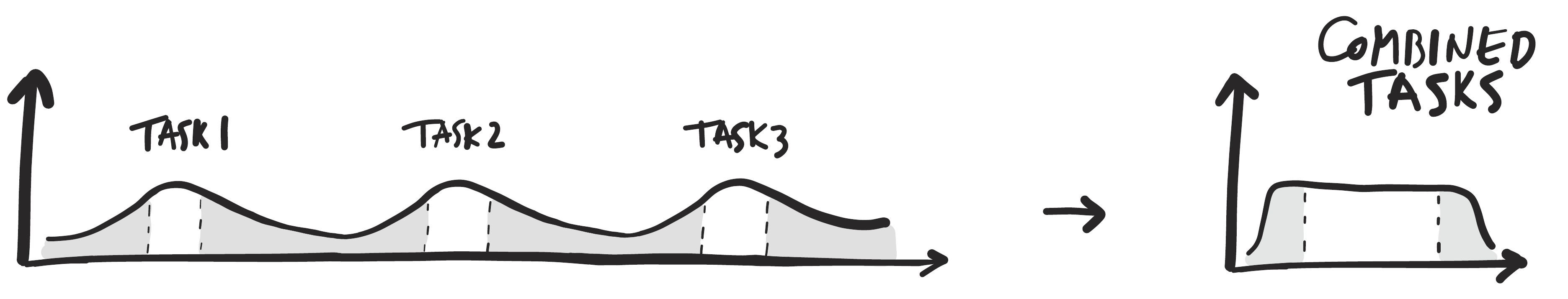}
  \caption{Illustration of the energy pattern Batch Operation. Energy usage can be reduced by combining the execution of different tasks.}
  \label{fig:batch_operations}
\end{figure}

\energyPattern{

Executing operations separately leads to extraneous tail energy consumptions~\citep{li2014investigation,corral2015energy,cai2015delaydroid}. As
illustrated in Figure~\ref{fig:batch_operations}, executing a task often induces
tail energy consumptions related with starting and stopping resources (e.g.,
starting a cellular connection).

}{

Bundle multiple operations in a single one. By combining multiple tasks, tail
energy consumptions can be optimized. Although background tasks can be
expensive, very often they have flexible time constraints. I.e., a given task
needs to be eventually executed, but it does not need to be executed in a
specific time.

}{

Use Job Scheduling APIs (e.g., `android.app.job.JobScheduler', `Firebase
JobDispatcher') that manage multiple background tasks occurring in a device.
These APIs will guarantee that the device will exit sleep mode only when there is a
reasonable amount of work to do or when a given task is urgent. It combines
several multiple tasks to prevent the device from constantly exiting sleep mode
(or doze mode). Other examples: execute low priority tasks only if another task
is using the same required resources; try to collect location data when other
apps are collecting it as well.

}

\subsection{Cache}

Avoid performing unnecessary operations by using cache mechanisms.

\energyPattern{

Typically mobile apps present data to users that is collect from a remote
server. However, it may happen that the same data is being collected from the
server multiple times.

}{

Implement caching mechanisms to temporarily store data from
a server~\citep{gottschalk2014saving}. In addition, verify whether there is an
update before downloading all data.

}{

Considering a social network app that shows other users' profiles. Instead of
downloading basic information and profile pictures every time a given profile is
opened, the app can use data that was locally stored from earlier visits.

}

\subsection{Decrease Rate}

Increase time between syncs/sensor reads as much as possible.

\energyPattern{

Mobile apps have to periodically perform operations. If the time between two
executions is small, the app will be executing operations more often. In some
cases, even if operations are executed more often, it will not affect users'
perception.

}{

Increase the delay between operations to find the minimal interval that does not
compromise user experience. This delay can be manually tuned by developers or
defined by users. More sophisticated solutions can also use context (e.g., time
of the day, history data, etc.) to infer the optimal update rate.

}{

Consider a news app that collects news from different sources, each one having
its own thread. Some news sources might have new content only once a week, while
others might be updated every hour. Instead of updating all threads at the same
rate, use data from previous updates to infer the optimal update rate of these
threads. Connect to the news source only if new updates are expected.
}

\subsection{User Knows Best}

Allow users to enable/disable certain features in order to save energy.

\energyPattern{

Energy efficiency solutions often provide a tradeoff between features and power
consumption. However, this tradeoff is different for different users --- some
users might be okay with fewer features but better energy efficiency, and vice
versa.

}{

Allow users to customize their preferences regarding energy critical
features. Since this might be more intuitive for power users, mobile apps
should provide optimal preferences by default for regular users.

}{

Consider a mail client for POP3 accounts as an example. In some cases, users are
not expecting any urgent message and are okay with checking for new mail in no
less than 10 minutes for the sake of energy efficiency. On the other hand, there
are cases in which users are waiting for urgent messages and would like to check
for messages every two minutes. Since there is no automatic mechanism to infer
the optimal update interval, the best option is to allow users to define it.

}

\subsection{Inform Users}

Let the user know if the app is doing any battery intensive operation.

\energyPattern{

There are specific use cases in mobile apps that can be energy greedy. On the
other hand, some features might be dropping user experience in order to improve
energy efficiency. If users do not know what to expect from the mobile app,
they might think it is not behaving correctly.

}{

Let users know about battery intensive operations or energy management
features. Properly flag this information in the user interface (e.g., alerts).

}{

Alert users when a power saving mode is active, or alert when a battery
intensive operation is about to be executed.

}

\subsection{Enough Resolution}

Collect or provide high accuracy data only when strictly necessary.

\energyPattern{

When collecting or displaying data, it is tempting to use high resolutions. The
problem of using data with high resolution is that its collection and
manipulation require more resources (e.g., memory, processing capacity, etc.).
As a consequence, energy consumption increases unnecessarily.

}{

For every use case, find the optimal resolution value that is required to
provide the intended user experience.

}{

Consider a running app that is able to record running sessions. While the user
is running, the app presents the current overall distance in real-time. While
calculating the most accurate value of the total distance would provide the
correct information, it would require precise real-time processing of GPS or
accelerometer sensors, which can be energy greedy. Instead, a lightweight method
could be used to estimate this information with lower but reasonable accuracy.
At the end of the session, the accurate results would still be processed, but
without real-time constraints.

}

\subsection{Sensor Fusion}

Use data from low power sensors to infer whether new data needs to be collected
from high power sensors

\energyPattern{

Mobile apps provide features that require reading data or executing operations
in different sensors or components. Such operations can be energy greedy,
causing high power consumption. Thus, they should be called as fewer times as
possible.

}{

Use complementary data from low power sensors to assure whether a given
energy-greedy operation needs to be executed.

}{

Use the accelerometer to infer whether the user has changed location. In the case
that the user is in the same location, data from GPS does not need to be
updated.

}

\subsection{Kill Abnormal Tasks}

Provide means of interrupting energy greedy operations (e.g., using timeouts, or
users input).

\energyPattern{

Mobile apps might feature operations that can be unexpectedly energy greedy
(e.g., taking a long time to execute).

}{

Provide a reasonable timeout for energy greedy tasks or wake locks.
Alternatively, provide an intuitive way of interrupting those tasks.

}{

In a mobile app that features an alarm clock, set a reasonable timeout for the duration of the
alarm. In case the user is not able to turn it off it will not drain the battery.

}

\subsection{No Screen Interaction}

Whenever possible allow interaction without using the display.

\energyPattern{

There are apps that require a continuous usage of the screen. However, there are
use cases in which the screen can be replaced by less power intensive
alternatives.

}{

Allow users to interact with the app using alternative interfaces (e.g., audio).

}{

In a navigation app, there are use cases in which users might be only using
audio instructions and do not need the screen to be on all the time. This
pattern is commonly adopted by audio players that use the earphone buttons
to play/pause or skip songs.

}

\subsection{Avoid Extraneous Graphics and Animations}

Graphics and animations are really important to improve the user experience.
However, they can also be battery intensive --- use them with
moderation~\citep{kim2016content}. This is also a recommendation in the
official documentation for iOS developers\footnote{\emph{Energy Efficiency
Guide for iOS Apps – Avoid Extraneous Graphics and Animations} available here:
\url{https://developer.apple.com/library/archive/documentation/Performance/Conce
ptual/EnergyGuide-iOS/AvoidExtraneousGraphicsAndAnimations.html} (Visited on
\today).}

\energyPattern{

Mobile apps often feature impressive graphics and animations. However, they need to be properly
tuned in order to prevent battery drain of users' devices. This is particularly critical in
e-paper devices.

}{

Study the importance of graphics and animations to the user experience. The
improvement in user experience may not be sufficient to cover the overhead on
energy consumption. Avoid using graphics animations or high-quality graphics.
Resort to low frame rates for animations when possible.

}{

For example, a high frame rate may make sense during game play, but a lower
frame rate may be sufficient for a menu screen. Use a high frame rate only when
the user experience calls for it.

}

\subsection{Manual Sync, On Demand}

Perform tasks exclusively when requested by the user.

\energyPattern{

Some tasks can be energy intensive, but not strictly necessary for some use cases
of the app.

}{

Provide a mechanism in the UI (e.g., button) that allows users to trigger energy
intensive tasks.

}{

In a beacon monitoring app, there are occasions in which the user does not need
to keep track of her/his beacons. Allow the user to start and stop monitoring
manually.

}

\section{Data Summary and Discussion}
\label{sec:data_summary}

In this section, we report and discuss findings regarding the presence of
energy patterns in the studied mobile applications. In particular, we study
differences between Android and iOS platforms and how often energy patterns
co-occur within the same app.

\subsection{Energy Patterns: Android vs. iOS}
\label{sec:android_vs_ios}

Next, we assess whether energy efficiency is addressed in a
different way in Android and iOS environments.

\subsubsection{Energy Efficiency Changes Per mobile app}

From the 1783 apps used in this study, we have found 332 ($19\%$) with at least one
commit, issue, or pull request related to energy efficiency. In Android we
have found 256 out of the 1021 apps ($25\%$), while in iOS we have found 76 out
of 756 apps ($10\%$). Congruent results are observed in the extraction of energy
patterns: we were able to extract energy patterns in 133 Android apps ($13\%$)
and 28 iOS apps ($4\%$). In general, Android developers put more effort into
improving the energy efficiency of their mobile apps than iOS developers.

However, research is necessary to explain why this is the case. First, our data
comprises only development activities that consciously address energy
efficiency. We understand that there are many factors that can affect these
results: power management mechanisms implemented by the system, documentation
of the frameworks, differences in targeted users, differences in targeted
devices, etc. For instance, the iOS platform provides APIs with
more strict power management rules, and developers are enforced to use
energy best practices beforehand even though they were not addressing energy
efficiency per se.

Developers even express their concern on not having their apps accepted in the
iOS App store for certain practices such as having tasks periodically running
in background\footnote{An example of developers dealing with the strict policy
of the iOS app store: \url{https://github.com/owncloud/ios/issues/13} (Visited
on \today).}. Although in this case the main goal is having the app accepted in the app
store, energy efficiency is addressed indirectly. On the contrary, Google Play
store, the official store for Android apps, is known to have less strict
policies~\citep{cuadrado2012mobile}.

\subsubsection{Prevalence of Patterns in Android and iOS}

We compare the prevalence of each energy pattern in the two platforms, Android
and iOS. Since our app dataset comprises a different number of apps for the two
platforms, we use the ratio of the number of occurrences $N_p$ of a given pattern
$p$ divided by the total number of apps ($M_X$) studied for a given platform
$X$:

\begin{equation}
  R(p,X) = \frac{N_p}{M_X},\ X=\{\mathrm{iOS}, \mathrm{Android}\}
\end{equation}

\begin{figure}
  \centering
  \includegraphics[width=\textwidth]{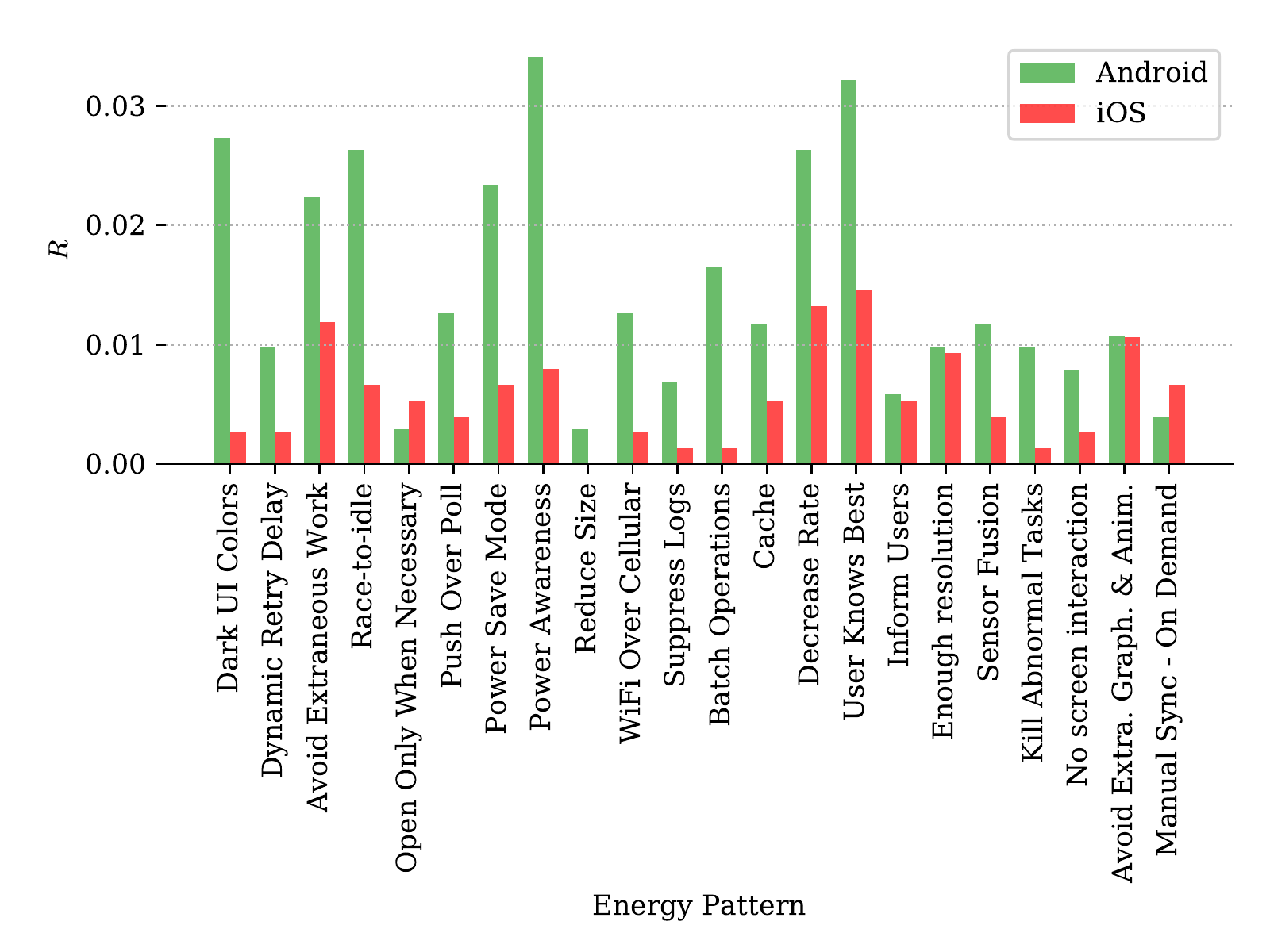}
  \caption{Comparison of the usage of energy patterns between Android and iOS mobile apps.}
  \label{fig:pattern_prevalence}
\end{figure}

Figure~\ref{fig:pattern_prevalence} shows the values of $R$ for every pattern
and platform. In general, energy patterns are more prevalent in Android rather
than iOS apps. Only two energy patterns were more frequent in iOS: \emph{Open
Only When Necessary} and \emph{Manual Sync, On Demand}. In addition, \emph{Inform Users},
\emph{Enough Resolution}, and \emph{Avoid Extraneous Graphics And Animations}
had a similar ratio of occurrences in both platforms. The remaining 18 patterns
were notably more frequent in Android apps.

The two platforms differ in the patterns that have the highest number of
occurrences.  In Android, the most frequent patterns were \emph{Power
Awareness}, \emph{User Knows Best}, \emph{Dark UI Colors}, and
\emph{Race-to-idle}. In iOS, the most frequent patterns were \emph{Avoid
Extraneous Work}, \emph{Decrease Rate}, \emph{User Knows Best}, and \emph{Avoid
Extraneous Graphics And Animations}. These patterns are being mentioned in the
official iOS documentation for developers, reinforcing the importance of
documentation to help developers build energy efficient
software~\citep{manotas2016empirical,sousa2018identifying}.

The pattern \emph{Dark UI Color} is considerably more popular amongst Android
apps than in iOS apps. This is an expected observation: only recently, iOS devices
started to feature AMOLED displays that reduce energy consumption when
using dark colors\footnote{Until mid 2018, iPhone X was the only iOS smartphone
with an AMOLED screen.}.

\subsection{Co-occurrence of Patterns}

\begin{figure}[h]
  \centering
    \includegraphics[width=\textwidth]{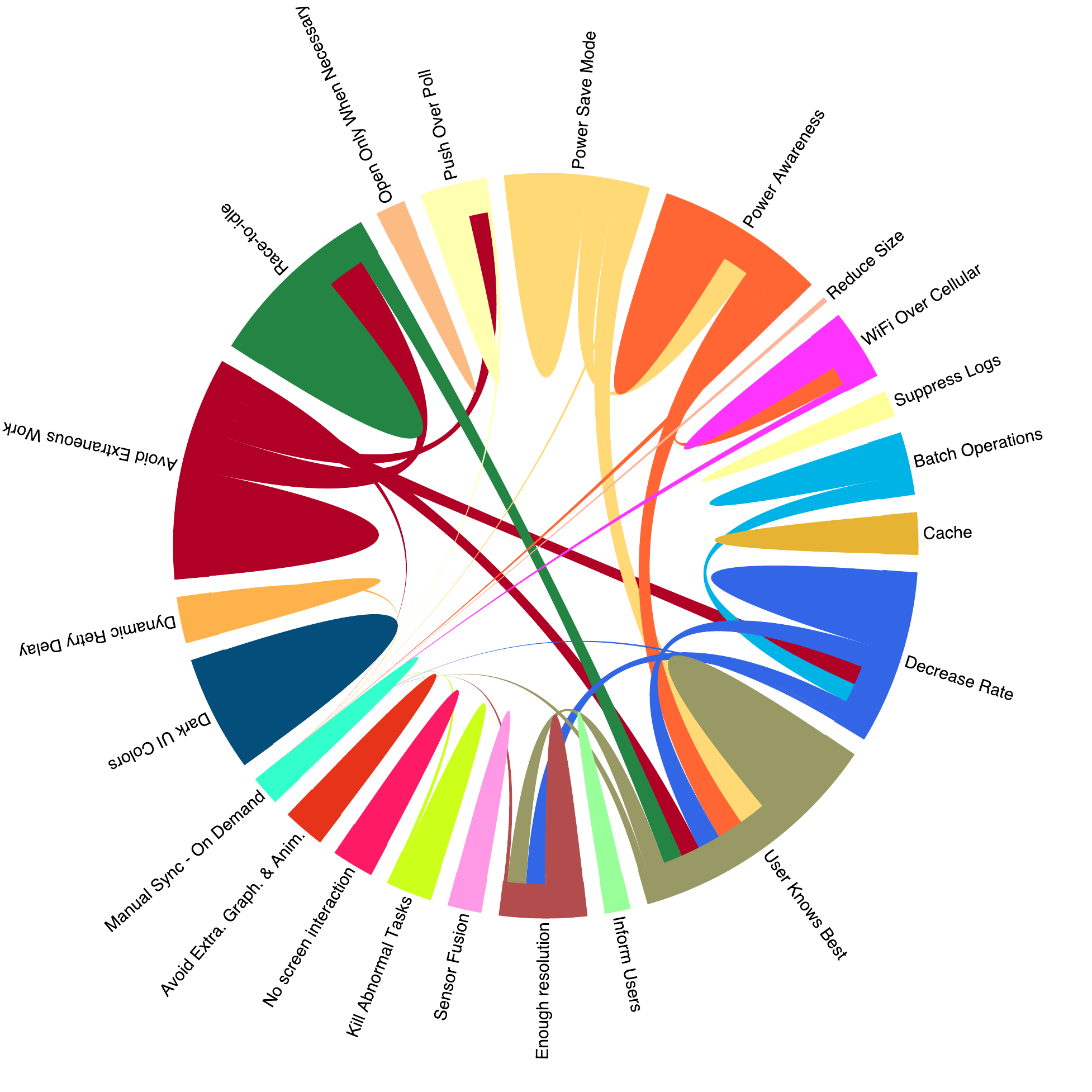}
  \caption{Co-occurrence of energy patterns in the same mobile application.}
  \label{fig:chord}
\end{figure}

We analyze which patterns tend to appear together within the
same mobile app. We resort to the chord diagram in Figure~\ref{fig:chord}. Each
pattern is connected to another pattern if found in the same app. The thicker
the edge the more frequent the pair of patterns co-occur.

To improve the interpretability of the chord diagram we have filtered out cases
in which two patterns co-occur less than 5 times. An interactive version
of the diagram containing all data is available in the online
catalog:~\url{https://tqrg.github.io/energy-patterns}.

The chord diagram of Figure~\ref{fig:chord} reveals the following relationships
between patterns:

\begin{itemize}
  \item \emph{Dark UI Colors} and \emph{Race-to-idle} are the most prevalent
    patterns (28 occurences), followed by \emph{Avoid Extraneous Work}
    (27 occurrences), \emph{User Knows Best} (25 occurences), and
    \emph{Decrease Rate} (24 occurences).

  \item \emph{User Knows Best} is a dominant pattern. It co-occurs with
  \emph{Power Save Mode}, \emph{Power Awareness},
  \emph{Decrease Rate}, \emph{Avoid Extraneous Work}, and \emph{Race To Idle}.

  \item \emph{Avoid Extraneous Work} is also a dominant pattern. It appears in
  apps that also use \emph{Race-to-Idle}, \emph{Push Over Poll},
  \emph{Decrease Rate}, and \emph{User Knows Best}.

  \item \emph{Power Save Mode}, \emph{Power Awareness}, and \emph{User Knows Best}
  are typically used together.
\end{itemize}

\subsection{Implications}

This catalog wraps up the techniques used by developers of open source mobile
apps to address energy efficiency. It helps developers understand how to design
energy efficient mobile apps by looking at solutions from other projects. Most
techniques are spread out in the literature, making energy efficiency a problem
that requires specialized developers. We mitigate this barrier by providing
common approaches to solve typical problems in the energy efficiency of mobile
apps.

Although we create energy patterns as a platform to share knowledge from
experienced developers, it can also serve as the base to future work on
automated tools to improve energy efficiency. Patterns such as \emph{Enough
resolution} may be challenging, requiring developers to have a deep
understanding of the devices they are targeting. Tools or APIs aiding to
address this issue would significantly decrease the efforts to adopt this
pattern. This is already the case for patterns such as \emph{Batch Operations}
that are featured in the platform's API: Android provides
the \texttt{JobScheduler} that schedules the jobs to execute at the most
efficient times. We understand that similar approaches should be leveraged to
other patterns. For instance, \emph{Power Save Mode} is a common pattern that
often requires re-implementing existing features.

We also find some remarks of interest to the research community. While some
techniques have been widely studied in previous work (e.g., \emph{Dark UI
Colors} and \emph{Race-to-idle}), many remain unnoticed. Patterns such as
\emph{Dynamic Retry Delay} or \emph{Kill Abnormal Tasks} can be argued based on
logical arguments. However, there is no empirical study that has evaluated the
cost and benefit of applying these patterns.

This catalog can also help educators include the topic of energy efficiency in
Mobile Software Engineering courses. Students can reach to this catalog to seek
more information on energy patterns and find examples of their application in real
Android and iOS apps.

\section{Threats to validity}
\label{sec:t2v}

In this section, we discuss potential systematic errors in our work and to what
extent our results can generalize.

\subsection{Internal validity}

We select all commits, issues and pull requests that contain the words
\emph{energy}, \emph{battery}, or \emph{power}. While we understand that this
covers the large majority of mentions of improvements in energy usage, some
mentions may have been missed. There are commits that may have been created to
improve energy efficiency, but the message may not mention it explicitly.
Moreover, we only use mentions written in English, which is the most common
language amongst developer communities.

In our methodology, we adopt a manual filtering of collected
commits, issues, and pull requests. While we understand that this was the safest
approach to avoid missing important mentions, human error can be expected from
this process. False positives may have been left in the dataset --- we argue
that during the thematic analysis these subjects are easily discarded. False
negatives (i.e., subjects of interest that were filtered out) may also occur due
to misinterpretation of subjects by authors during the selection. However, we expect
this to comprise a negligible number of cases.


Only commits merged with the \textit{default} branch of projects were considered.
Energy improvements that were implemented in different branches of the project
were not considered in our study. Although some interesting patterns may lie in
some of these branches we cannot guarantee their quality: changes that have not
been merged are still lacking the validation from the development team.

In addition, we only collect energy patterns that were used by mobile app
development practitioners and are available on open source projects. There
might be other patterns that improve energy efficiency but that are not being
used by the community. Those patterns are out of the scope of this study and
are left for future work. Moreover, patterns that also improve other properties
besides energy usage might occur in the projects in this study.

We solely list cases that occur with the main goal of improving energy
efficiency. In this study, we only address energy improvements that are clearly
described in the description of the respective commit, issue, or pull request.
However, the classification of developers' intent in code changes is a
non-trivial open problem~\citep{pascarella2018self}. Thus, less obvious energy
improvements were not studied. In addition, this catalog is based on
developers common approaches to address energy efficiency. We do not measure
the magnitude of potential gains of these patterns work. An empirical study
would help to assess these gains, as done in previous
work~\citep{carette2017investigating,palomba2019impact}.

Finally, we have discarded less significant
patterns by filtering out patterns with less than three occurrences. We
understand that some of them may hide interesting energy efficiency strategies.
However, no literature was found to support these patterns, and assessing their
impact is out of the scope of this paper.

\subsection{External validity}

The data analyzed in this work is typically private for commercial apps. Thus, we
focus exclusively on open source mobile apps. Development of commercial apps is
usually driven by different goals and budgets. Hence, energy usage may be
targeted in a different way in these apps. However, energy patterns collected in
this study can be applied in any mobile app project regardless of its license.

We only analyze iOS and Android mobile apps. While this comprises most of the
mobile apps in the market, other mobile platforms also have their share (Windows
Phone, BlackBerry OS, Firefox OS, Ubuntu Touch, etc.). We discarded coding
practices for energy efficiency that we understood being specific to the
platform in which they were implemented (e.g., using certain API methods).
Unless the mobile operative system or the hardware device deals with the
contexts identified in our work under the hood, we expect these patterns to be
useful in other platforms.

\section{Conclusion}
\label{sec:conclusions}

We analyzed commits, issues and pull requests from 1021 Android apps and 756
iOS apps to identify design practices to improve the energy efficiency of
mobile apps. As an outcome, this work delivers a catalog of 22 design patterns
based on 1563 energy-related changes of mobile apps. To the best of our
knowledge, this is the first time energy practices for mobile apps are studied
at a large scale in both iOS and Android. This catalog will help mobile app
designers and developers make educated decisions when building (energy
efficient) apps, regardless of the target platform.

We leverage a dataset with changes that address energy efficiency in mobile
apps for Android and iOS. In addition, we list the occurrence of energy
patterns in this dataset. The dataset is publicly available and we welcome
contributions from the community as pull request. An interesting remark from
this dataset is that it comprises only $19\%$ of the total number of mobile
apps that we analyze. This means that only a minority of mobile apps have had
changes to improve energy efficiency. Moreover, results show that efforts to
improve energy efficiency are more common within Android apps (25\%) than iOS
apps (10\%). Further research should investigate the root causes of this
observation.

As future work, it would be interesting to implement these patterns in an
automated refactoring tool. Similar tools have been delivered in previous work
on quality smells~\citep{palomba2017lightweight,cruz2018using}. Moreover, we
would like to study these patterns in the context of Cyber-Physical Systems and
Internet of Things applications, in which power consumption has been identified
as a challenge to be addressed~\citep{white2010r,palattella2016internet}.
Finally, we plan to continuously extend the catalog with a broader set of
energy patterns. We welcome contributions from the mobile app development
community as pull requests in our online repository.

\section*{Acknowledgment}
This work is financed by the ERDF --- European Regional Development Fund
through the Operational Program for Competitiveness and Internationalization -
COMPETE 2020 Program and by National Funds through the Portuguese funding
agency, FCT - Funda\c{c}\~ao para a Ci\^encia e a Tecnologia within project
POCI-01-0145-FEDER-016718. Luis Cruz is sponsored by an FCT scholarship grant
number PD/BD/52237/2013. Furthermore, we would like to thank Sofia Reis for the
illustrations of energy patterns.%
\begin{appendices}
\section{Grey Literature}
\label{sec:grey_literature}

\begin{enumerate}[label={[\arabic{enumi}]},ref=\arabic{enumi}]

  \item\label{grey:apple_graphics} Apple Developer Documentation Archive. Energy
  Efficiency Guide for iOS Apps – Avoid Extraneous Graphics and Animations.
  URL:~\url{https://developer.apple.com/library/archive/documentation/Performance/Conceptual/EnergyGuide-iOS/AvoidExtraneousGraphicsAndAnimations.html}

  \item\label{grey:xpp} Daniel Gultsch. The State of Mobile XPP in 2016. URL:~\url{https://gultsch.de/xmpp_2016.html}

  \item\label{grey:alternative_push} Alternative Push Notification Transport. URL:~\url{https://github.com/matrix-org/GSoC/blob/master/IDEAS.md#alternative-push-notification-transport}

  \item\label{grey:apple_power_save} Apple Developer Documention Archive. Energy Efficiency Guide for iOS Apps – React to Low Power Mode on iPhones. URL:~\url{https://developer.apple.com/library/archive/documentation/Performance/Conceptual/EnergyGuide-iOS/LowPowerMode.html}

  \item\label{grey:gzip} Stackoverflow. Is gzip compression useful for mobile devices?. URL:~\url{https://stackoverflow.com/questions/3065920/is-gzip-compression-useful-for-mobile-devices}

  \item\label{grey:wifi_cellular} Mitch Bartlett. Does Wi-Fi Consume More Battery Power Than 3G or 4G/LTE?. URL:~\url{https://www.technipages.com/does-wi-fi-consume-more-battery-power-than-3g-or-4glte}

  \item\label{grey:android_connectivity} Android SDK Documentation. Modifying your Download Patterns Based on the Connectivity Type. URL:~\url{https://developer.android.com/training/efficient-downloads/connectivity_patterns}

  \item\label{grey:apple_network} Apple Developer Documentation Archive. Energy Efficiency Guide for iOS Apps – Energy and Networking. URL:~\url{https://developer.apple.com/library/archive/documentation/Performance/Conceptual/EnergyGuide-iOS/EnergyandNetworking.html}

  \item\label{grey:android_doze} Android SDK Documentation. Optimizing for Doze and App Standby. URL:~\url{https://developer.android.com/training/monitoring-device-state/doze-standby}

  \item\label{grey:android_battery} Android SDK Documentation. Android Developer Guides — Optimizing for Battery Life. URL:~\url{https://developer.android.com/topic/performance/power/}

  \item\label{grey:android_youtube} Android Developers Youtube Channel. DevBytes: Efficient Data Transfers - Batching, Bundling, and SyncAdapters. URL:~\url{https://www.youtube.com/watch?v=5onKZcJyJwI}

  \item\label{grey:apple_core} Apple's Core Location Documentation. CLLocationManager. URL:~\url{https://developer.apple.com/documentation/corelocation/cllocationmanager}

\end{enumerate}
\end{appendices}
\balance
\bibliographystyle{IEEEtranN}
\bibliography{bibliography}
\end{document}